\documentclass[a4paper,11pt]{article}
\usepackage[affil-it]{authblk} 
\usepackage{hyperref} 
\usepackage{enumitem}
\usepackage{amssymb}
\usepackage{amsmath}
\usepackage{multirow}
\providecommand{\keywords}[1]
{
    \small    
    \textbf{\textit{Keywords---}} #1
}

\newcommand{\ket}[1]{\ensuremath{\left|#1\right\rangle}}

\newtheorem{example}{\bf Example}
\begin{document}
    
    \title{Cryptanalysis of Quantum Secure Direct Communication Protocol with Mutual Authentication Based on Single Photons and Bell States}
    
    \author{Nayana Das%
        \thanks{Email address: \texttt{dasnayana92@gmail.com} }}
    \affil{Applied Statistics Unit, Indian Statistical Institute, Kolkata, India.}
    
    \author{Goutam Paul %
        \thanks{Email address: \texttt{goutam.paul@isical.ac.in}}}
    \affil{Cryptology and Security Research Unit, R. C. Bose Centre for Cryptology and Security, Indian Statistical Institute, Kolkata, India.\\}
    
    \date{}
    
    \maketitle
    
    \begin{abstract}
        Recently, Yan et al. proposed a quantum secure direct communication (QSDC) protocol with authentication using single photons and Einstein-Podolsky-Rosen (EPR) pairs (Yan et al., CMC-Computers, Materials \& Continua, 63(3), 2020). In this work, we show that the QSDC protocol is not secure against intercept-and-resend attack and impersonation attack. An eavesdropper can get the full secret message by applying these attacks. We propose a modification of this protocol, which defeats the above attacks along with all the familiar attacks.
    \end{abstract}
    
    
    \section{Introduction}
    \label{intro}
    Quantum cryptography is an application of quantum mechanics in the field of cryptography, which provides unconditional security based on the laws of physics. In 1984, Bennett and Brassard proposed the first quantum cryptographic protocol, which is a quantum key distribution (QKD) protocol, also called the BB84 QKD~\cite{bennett2020quantum}. Since then various types of QKD protocols have been proposed, such as QKD with entanglement~\cite{ekert1991quantum,long2002theoretically,li2016one}, without entanglement~\cite{bennett1992quantum1,lucamarini2005secure}, experimental QKD~\cite{bennett1992experimental,zhao2006experimental,tang2014experimental,bedington2016nanosatellite,zhong2019proof} and so on.
    
    Quantum secure direct communication (QSDC) is another direction of quantum cryptography, which offers secure communication without any shared key~\cite{long2002theoretically,beige2001secure,bostrom2002ping,deng2003two,deng2004secure,wang2005quantum,hu2016experimental,zhang2017quantum}. In QSDC protocols, the sender encodes the secret message into some qubits by using some predefined encoding rules and sends those qubits to the receiver. After some security checks, the receiver can get back the secret message. Some interesting generalization of QSDC protocols are quantum dialogue or bidirectional QSDC~\cite{nguyen2004quantum,zhang2004deterministic,zhong2005quantum,xia2006quantum,xin2006secure,gao2010two,das2020two}, multi-party QSDC~\cite{gao2005deterministic,jin2006three,ting2005simultaneous,tan2014multi} and so on. 
    
    If QSDC or any quantum cryptographic protocol is not properly designed, it gives a chance to an eavesdropper to impersonate an authorized party. For this concern, each legitimate party should verify the authenticity of other parties, which requires quantum authentication protocols~\cite{curty2001quantum,shi2001quantum,lee2005quantum}. The first QSDC protocol with authentication was proposed in 2006~\cite{lee2006quantum}, and thereafter many researchers are working in this domain~\cite{dan2010new,chang2014controlled,hwang2014quantum}. 
    
    There are multiple quantum cryptographic protocols, which are proven to be insecure against various familiar  attacks, such as, intercept-and-resend attack~\cite{qin2010improving,wei2011comment,chang2015intercept}, impersonation attack~\cite{zhang2007comment,su2009high,fei2011cryptanalysis}, Denial-of-Service attack~\cite{cai2004ping,yang2012comment,liu2014cryptanalysis}, man-in-the-middle attack~\cite{peev2005novel,lin2013man}, entangle-measure attack~\cite{liu2014cryptanalysis,fei2008special}, Trojan horse attack~\cite{deng2005improving,gisin2006trojan} etcetera. These are all active attacks, i.e., an eavesdropper has access to the communicated qubits in the quantum channel between the legitimate parties, and actively participates in the protocol. Some inactive attack also causes information leakage problems in some communication protocols~\cite{gao2008comment,das2020improving}. 
    
    In 2020, Yan et al. have presented a QSDC protocol based on single photons and EPR pairs, which also realizes the mutual authentication~\cite{Yan.2020.cmc}. For simplicity, throughout this paper, we call this QSDC protocol as YZCSS protocol. In this protocol, Alice, the message sender, prepares qubit pairs corresponding to the secret message and her authentication identity. She sends all the qubits to Bob, the message receiver, who uses his authentication identity to recover the secret message. However, in this article, we show that the YZCSS protocol is not secure against intercept-and-resend attack and impersonation attack. If an eavesdropper applies any one of these attacks, then it can get the complete secret message, i.e., not only a portion of the message is revealed, but also the entire message is compromised. Moreover, for impersonation attack, the legitimate parties can not realize the presence of the eavesdropper. Furthermore, we present a modification of the YZCSS protocol to improve its security.
    
    The rest of the paper is organized as follows: in Section~\ref{sec:review}, we briefly describe the YZCSS protocol, then in the next section we discuss the security flaws of the YZCSS protocol. An improved version of the protocol is presented in Section~\ref{sec4:modified protocol} and finally we conclude our result.

    \section{Brief review of the YZCSS protocol}
    \label{sec:review}
    In this section, we describe the YZCSS protocol. There are two parties, namely, Alice and Bob with their corresponding identities $ID_A$ and $ID_B$ respectively, where $ID_A,~ID_B \in \{0,1\}^N$. Alice wants to send a secret message $M\in \{0,1\}^N$ to Bob by using single photons and Bell states, where the Bell states (EPR pairs) are defined as: 
    \begin{equation}
    \ket{\Phi^{\pm}}=\frac{1}{\sqrt{2}}(\ket{00} \pm \ket{11}),\\
    \ket{\Psi^{\pm}}=\frac{1}{\sqrt{2}}(\ket{01} \pm \ket{10}).
    \end{equation}
    The steps of the protocol are as follows:
    \begin{enumerate}
        \item Alice and Bob have their previously shared identities $ID_A$ and $ID_B$, they used some QKD to exchange $ID_A$ and $ID_B$. Alice prepares two ordered sets of two-qubit states $S_M$ and $S_A$ corresponding to the message $M$ and her own identity $ID_A$, each ordered set contains $N$ qubit pairs.
        For $1 \leq i \leq N$, let the $i$-th bit of $M$ (or $ID_A$ or $ID_B$) be $M_i$ (or $ID_{A,i}$ or $ID_{B,i}$) and the $i$-th qubit of $S_M$ (or $S_A$) be $S_{M,i}$ (or $S_{A,i}$). She prepares the qubits by using the following rule:
        \begin{enumerate}
            \item if $M_i$ (or $ID_{A,i}$) $=0$, then $S_{M,i}$ (or $S_{A,i}$) $=\ket{01}$ or $\ket{10}$ with equal probability,
            \item if $M_i$ (or $ID_{A,i}$) $=1$, then $S_{M,i}$ (or $S_{A,i}$) $=\ket{\Phi^{+}}$ or $\ket{\Phi^{-}}$ with equal probability.
        \end{enumerate}
        The qubit pairs of the ordered set $S_A$ are called decoy states. Now Alice inserts these decoy states into the ordered set $S_M$ according to the following rule: 
        \begin{enumerate}
            \item if $ID_{B,i}=0$, then she inserts $S_{A,i}$ before $S_{M,i}$, and
            \item if $ID_{B,i}=1$, then she inserts $S_{A,i}$ after $S_{M,i}$.
        \end{enumerate}
        Let the new ordered set be $S$ containing $2N$ qubit pairs. Then Alice sends $S$ to bob using a quantum channel. Let us take an example. 
        
        \begin{example} Let $M=10110$, $ID_{A}=01101$ and $ID_{B}=01001$.\\
            Then $S_M=\{\ket{\Phi^{+}},\ket{01},\ket{\Phi^{+}},\ket{\Phi^{-}},\ket{01}\}$,
            $S_A=\{\ket{10},\ket{\Phi^{-}},\ket{\Phi^{-}},\ket{01},\ket{\Phi^{+}}\}$ and \\
            $S=\{\ket{10},\ket{\Phi^{+}},\ket{01},\ket{\Phi^{-}},\ket{\Phi^{-}},\ket{\Phi^{+}},\ket{01},\ket{\Phi^{-}},\ket{01},\ket{\Phi^{+}}\}$.
        \end{example}
        
        \item After Bob receives $S$, he knows the exact positions of the decoy photons corresponding to his identity $ID_B$. Bob measures those decoy photons in proper bases according to $ID_A$. If $ID_{A,i}=0$, then he chooses $Z\times Z$ basis, where $Z=\{\ket{0},\ket{1}\}$, thus $Z\times Z=\{\ket{00},\ket{01},\ket{10},\ket{11}\}$, and if  $ID_{A,i}=1$, then he chooses the Bell basis  $=\{\ket{\Phi^{+}},\ket{\Phi^{-}},\ket{\Psi^{+}},\ket{\Psi^{-}}\}$ to measure $S_{A,i}$.
        Bob also measures the qubit pairs of $S_M$ in $Z\times Z$ basis or Bell basis randomly. He notes the measurement results.
        \item Bob asks Alice to announce the initial states of the qubit pairs of $S_A$ for security check. They compare the initial states and the measurement results of the decoy photons and calculate the error rate. If the error rate exceeds some pre-defined threshold value, then they terminate the protocol, else they continue.
        \item Bob gets all the secret message bits from the measurement results of the qubit pairs of $S_M$. The relation between the measurement results and the secret message bits are given in Table~\ref{table-1}.
        To check the integrity of the secret message Alice and Bob publicly compare some parts of the message. 
        
    \end{enumerate}
    The authors of~\cite{Yan.2020.cmc} have shown that the YZCSS protocol is secure against various kinds of attacks, such as impersonation attack, intercept-and-resend attack, man-in-the-middle attack, entangle-measure attack. However, in the next section, we show that an eavesdropper can design a strategy that allows him to effectively execute the intercept-and-resend attack. A similar argument follows for impersonation attack as well, making this protocol insecure against these two attacks.
    
    \begin{table}[]
        \centering
        \caption{Different cases of the YZCSS protocol}
        \label{table-1}
        \begin{tabular}{|c|c|c|c|c|}
            \hline
            \textbf{Secret message} & \textbf{Encoded}   & \textbf{Basis chosen} & \textbf{measurement}                                   & \textbf{Decoded } \\
            \textbf{ bit of Alice} {$\mathbf{M_i}$} & \textbf{qubit} $\mathbf{S_{M,i}}$   & \textbf{by Bob} & \textbf{result of Bob}                                   & \textbf{secret bit} \\ \hline
            \multirow{4}{*}{0}                        & \multirow{2}{*}{$\ket{01}$}        & $Z\times Z$ basis                                & $\ket{01}$                             & 0                           \\ \cline{3-5} 
            &                                    & Bell basis                               & $\ket{\Psi^{+} }$ or $\ket{\Psi^{-} }$ & 0                           \\ \cline{2-5} 
            & \multirow{2}{*}{$\ket{10}$}        & $Z\times Z$ basis                                & $\ket{10}$                             & 0                           \\ \cline{3-5} 
            &                                    & Bell basis                               & $\ket{\Psi^{+} }$ or $\ket{\Psi^{-} }$ & 0                           \\ \hline
            \multirow{4}{*}{1}                        & \multirow{2}{*}{$\ket{\Phi^{+} }$} & $Z\times Z$ basis                                & $\ket{00}$ or $\ket{11}$               & 1                           \\ \cline{3-5} 
            &                                    & Bell basis                               & $\ket{\Phi^{+} }$                      & 1                           \\ \cline{2-5} 
            & \multirow{2}{*}{$\ket{\Phi^{-} }$} & $Z\times Z$ basis                                & $\ket{00}$ or  $\ket{11}$              & 1                           \\ \cline{3-5} 
            &                                    & Bell basis                               & $\ket{\Phi^{-} }$                      & 1                           \\ \hline
        \end{tabular}
    \end{table}
    
    \section{Security loophole of the YZCSS protocol}\label{sec3:security}
    We now show that the YZCSS protocol discussed in the previous section is not secure against intercept-and-resend attack and impersonation attack, an eavesdropper ($Eve$) can get the whole secret message $M$ and Alice's authentication identity $ID_A$ by adopting these attacks.
    
    \subsection{Intercept-and-resend attack}
    In this attack strategy, when Alice sends the quantum states to Bob, $Eve$ intercepts those from the quantum channel, he measures the states and resends those to Bob. However, to attack the YZCSS protocol, $Eve$ follows a special strategy while resending the quantum states to Bob. The process of the attack is as follows.
    \begin{enumerate}
        \item $Eve$ intercepts the ordered set $S$ and measures each two-qubit state randomly in $Z\times Z$ basis or Bell basis and notes down the measurement results. For $1 \leq i \leq 2N$, if he chooses $Z\times Z$ basis to measure the $i$-th qubit pair of $S$ and the measurement result is either $\ket{01}$ or $\ket{10}$, then he simply sends this state to Bob. But if the measurement result is either $\ket{00}$ or $\ket{11}$, $Eve$ definitely knows that he chooses the wrong basis and the initial state was either $\ket{\Phi^{+}}$ or $\ket{\Phi^{-}}$. Then he randomly prepares $\ket{\Phi^{+}}$ or $\ket{\Phi^{-}}$ and sends it to Bob. Similarly if $Eve$ chooses Bell basis and gets $\ket{\Phi^{+}}$ or $\ket{\Phi^{-}}$, then sends them. Otherwise, he randomly sends $\ket{01}$ or $\ket{10}$ to Bob.
        
        \item $Eve$ constructs a $2N$-bit string $m$ from the measurement results by using Table~\ref{table-eve}.
        \begin{table}[h]
            \caption{Rule of construction of $m$ by $Eve$}
            \label{table-eve}
            \begin{tabular}{|c|c|c|}
                \hline
                \textbf{Basis chosen by $\mathbf{Eve}$} & \textbf{$\mathbf{Eve}$'s measurement result}    & \textbf{Corresponding bit of $\mathbf{m}$} \\ \hline
                \multirow{2}{*}{$Z\times Z$ basis}                 & $\ket{01}$ or $\ket{10}$               & 0                                 \\ \cline{2-3} 
                & $\ket{00}$ or $\ket{11}$               & 1                                 \\ \hline
                \multirow{2}{*}{Bell basis}                & $\ket{\Psi^{+} }$ or $\ket{\Psi^{-} }$ & 0                                 \\ \cline{2-3} 
                & $\ket{\Phi^{+} }$ or $\ket{\Phi^{-} }$ & 1                                 \\ \hline
            \end{tabular}
        \end{table}
        
        \item \label{eve-msg} $Eve$ splits the $2N$-bit string $m=m_1 m_2 \ldots m_{2N}$ into $N$ number of $2$-bit strings $\mathcal{M}_1,\mathcal{M}_2,\ldots ,\mathcal{M}_N$, and for $1 \leq i \leq N$, $\mathcal{M}_i=m_{2i-1}m_{2i}$. Now from the construction procedure of the ordered set $S$, $Eve$ exactly knows that each $\mathcal{M}_i$ contains the $i$-th bit of secret message $M$ and the $i$-th bit of Alice's authentication identity $ID_A$. If both the bits of $\mathcal{M}_i$ are equal, i.e., $\mathcal{M}_i=bb$, where $b \in \{0,1\}$, then he concludes $M_i=b$ and $ID_{A,i}=b$. Again if $\mathcal{M}_i=b\bar{b}$, where $\bar{b}$ = bit complement of $b$, then he waits for Alice's announcement about the initial states of the decoy photons. If she announces $\ket{01}$ or $\ket{10}$, then $Eve$ concludes $ID_{A,i}=0$ and $M_i=1$, otherwise he concludes $ID_{A,i}=1$ and $M_i=0$. Thus $Eve$ can successfully attack the protocol and gets the complete secret message.
    \end{enumerate}
    Now Alice and Bob can detect this intercept-and-resend attack at the time of security check, but it has no impact on the attack result as one of the main requirement of a QSDC protocol is: ``the secret messages which have been encoded already in the quantum states should not leak even though an eavesdropper may get hold of channel"~\cite{deng2003two}. 
    
    \subsection{Impersonation attack}
    By analyzing the YZCSS protocol, we find that the authentication procedure of this QSDC protocol is unidirectional, i.e., only Bob can verify Alice's identity.  
    Here we show that how $Eve$ impersonate Bob to acquire the secret message of Alice. The process is as follows:
    \begin{enumerate}
        \item Alice prepares the ordered set $S$ and sends it to $Eve$.
        \item After receiving $S$, $Eve$ measures all the qubit pairs randomly in $Z\times Z$ or Bell basis and generates a $2N$-bit string $m$ from the measurement results by using Table~\ref{table-eve}.
        \item $Eve$ asks Alice to declare the initial state of the decoy photons and from this information, he gets the whole secret message (by using the same process as in Step~\ref{eve-msg} of the intercept-and-resend attack).   
    \end{enumerate}
    In this case, Alice can not detect $Eve$, or in other words, only one-way authentication is possible in the YZCSS protocol. Moreover, without knowing the exact position of the decoy photons, $Eve$ can get the whole secret message.
    
    Let us take an example of this attack.
    \begin{example}
        Let $M=10110$, $ID_{A}=01101$ and $ID_{B}=01001$.\\
        Then $S_M=\{\ket{\Phi^{+}},\ket{01},\ket{\Phi^{+}},\ket{\Phi^{-}},\ket{01}\}$,
        $S_A=\{\ket{10},\ket{\Phi^{-}},\ket{\Phi^{-}},\ket{01},\ket{\Phi^{+}}\}$ and \\
        $S=\{\ket{10},\ket{\Phi^{+}},\ket{01},\ket{\Phi^{-}},\ket{\Phi^{-}},\ket{\Phi^{+}},\ket{01},\ket{\Phi^{-}},\ket{01},\ket{\Phi^{+}}\}$.
        \begin{enumerate}
            \item $Eve$ has the ordered set $S$.
            \item Let  $\mathcal{B}=\{Z,Z,\text{Bell},Z,\text{Bell},\text{Bell},\text{Bell},Z,Z,\text{Bell}\}$ be a sequence of bases which $Eve$ choses to measure the qubit pairs of $S$.
            \item Let the ordered set of measurement results be\\ $\{\ket{10},\ket{00},\ket{\Psi^{-}},\ket{11},\ket{\Phi^{-}},\ket{\Phi^{+}},\ket{\Psi^{+}},\ket{11},\ket{01},\ket{\Phi^{+}}\}$. 
            \item Then $m=0101110101$ and $\mathcal{M}_1=01$, $\mathcal{M}_2=01$, $\mathcal{M}_3=11$, $\mathcal{M}_4=01$, $\mathcal{M}_5=01$. $Eve$ concludes $M_3=1$ and $ID_{A,3}=1$.
            \item Alice announces $S_A=\{\ket{10},\ket{\Phi^{-}},\ket{\Phi^{-}},\ket{01},\ket{\Phi^{+}}\}$ and then $Eve$ concludes 
            \begin{itemize}
                \item $ID_{A,1}=0$ and $M_1=1$,
                \item $ID_{A,2}=1$ and $M_2=0$,
                \item $ID_{A,4}=0$ and $M_4=1$,
                \item $ID_{A,5}=1$ and $M_5=0$.
            \end{itemize} 
            Thus $Eve$ gets the whole secret message $M=10110$.
        \end{enumerate}
    \end{example}
    Another problem of the YZCSS protocol is that the length of the authentication identities of Alice and Bob are equal to the length of the secret message. Since the identities are previously shared, Alice can send a fixed length message to Bob, which is a disadvantage of this protocol.
    In the next section, we propose a remedy to these security problems of the YZCSS protocol.
    
    \section{Proposed modification }\label{sec4:modified protocol}
    Now we discuss how to modify this YZCSS protocol so that it can provide mutual authentication and stand against the intercept-and-resend attack. In the original protocol, the length of $ID_A$ and $ID_B$ are equal to the length of the message, which may vary. However, in our improved version, we fix the length of $ID_A$ and $ID_B$, and the fixed-length is unknown to any third party. Here we use some techniques of the authentication protocol proposed by Fei et al.~\cite{fei2011cryptanalysis}. Our modified protocol is given below:
    \begin{enumerate}
        \item Qubits preparation to encode secret message:\label{mod-qubit}
        \begin{enumerate}
            \item Alice and Bob have their previously shared $k$-bit identities $ID_A$ and $ID_B$, where $k$, $ID_A$ and $ID_B$ are unknown to everybody other than Alice and Bob. Alice prepares an ordered set of $N$ qubit pairs $S_M$ corresponding to her $N$-bit message $M$.
            For $1 \leq i \leq N$, she prepares the qubit pairs of $S_M$ by using the following rule:
            
            \begin{equation}
            M_i=
            \begin{cases}
            0 \Rightarrow S_{M,i} =\ket{01}$ \text{or} $\ket{10} \text{, with equal probability;}\\
            1 \Rightarrow S_{M,i} =\ket{\Phi^{+}}$ \text{or} $\ket{\Phi^{-}} \text{, with equal probability.}
            \end{cases}
            \end{equation}
            
            She applies a random permutation on the ordered set $S_M$ containing $2N$ qubits and let the new ordered set be $Q_M$. For $1 \leq j \leq 2N$, let the $j$-th qubit $Q_M$ be $Q_{M,j}$. Note that the two qubits of each qubit pair corresponding to the message bits are in two random positions of $Q_M$.
            \item Alice prepares the first ordered set of decoy photons $S_A$, for authentication, corresponding to her own identity $ID_A$ as follows: for $1 \leq i \leq k$,
            \begin{equation}
            ID_{A,i}=
            \begin{cases}
            0 \Rightarrow S_{A,i} =\ket{0}$ \text{or} $\ket{1} \text{, with equal probability;}\\
            1 \Rightarrow S_{A,i} =\ket{+}$ \text{or} $\ket{-} \text{, with equal probability,}
            \end{cases}
            \end{equation}
            where $\ket{+}=\frac{1}{\sqrt{2}}(\ket{0}+\ket{1})$ and $\ket{-}=\frac{1}{\sqrt{2}}(\ket{0}-\ket{1})$. Now she inserts these decoy states into the ordered set $Q_M$ according to the following rule: for $1 \leq i \leq k$,
            \begin{enumerate}
                \item if $ID_{B,i}=0$, then she inserts $S_{A,i}$ before $Q_{M,\lambda i-\lambda-1}$,
                \item if $ID_{B,i}=1$, then she inserts $S_{A,i}$ after $Q_{M,\lambda i}$,
            \end{enumerate}
            where $\lambda=[2N/k]$, $[x]=$ greatest integer not greater than $x$ and $k \leq N$.
            Let the new ordered set be $S$ containing $2N+k$ qubits. For better understanding, let us take an example,
            \begin{example}
                Let $M=1011010$, $ID_{A}=011$ and $ID_{B}=010$.
                \begin{enumerate}
                    \item $S_M=\{\ket{\Phi^{+}},\ket{01},\ket{\Phi^{+}},\ket{\Phi^{-}},\ket{01},\ket{\Phi^{+}},\ket{10}\}$ and let the $i$-th pair of $S_M$ be $(S_{M,i}^1,S_{M,i}^2)$.
                    \item $Q_M= \{S_{M,6}^1,S_{M,4}^2, S_{M,2}^2, S_{M,4}^1, S_{M,3}^2, S_{M,1}^1, S_{M,2}^1,S_{M,7}^1, S_{M,1}^2, S_{M,5}^1,S_{M,6}^2, S_{M,3}^1, S_{M,5}^2, S_{M,7}^2\}$.
                    \item $S_A=\{\ket{0},\ket{-},\ket{-}\}$.
                    \item $\lambda=[14/3]=4$.
                    \item $S= \{\ket{0},S_{M,6}^1,S_{M,4}^2, S_{M,2}^2, S_{M,4}^1,
                     S_{M,3}^2, S_{M,1}^1, S_{M,2}^1,S_{M,7}^1, \ket{-},
                      \ket{-}, S_{M,1}^2, S_{M,5}^1,S_{M,6}^2, S_{M,3}^1,$\\ $
                       S_{M,5}^2, S_{M,7}^2\}$.
                \end{enumerate}     
            \end{example}
            \item She prepares a second set of decoy photons randomly from $\{\ket{0},\ket{1},\ket{+},\ket{-}\}$ and inserts them in random positions of $S$ and sends the new ordered set $S'$ to Bob using a quantum channel.
        \end{enumerate}
        \item \label{mod-security} Security check: After Bob receives $S'$, Alice announces the positions and bases of the second set of decoy photons. Bob measures those decoy photon and they calculate the error rate in the channel by comparing the measurement results with the initial states. If the error rate is low, then they continue the protocol, otherwise terminate this. 
        
        \item Authentication procedure:\label{mod-auth} 
        \begin{enumerate}
            \item Bob knows the exact positions of the decoy photons of $S_A$ corresponding to his identity $ID_B$. He measures those decoy photons in proper bases according to $ID_A$. If $ID_{A,i}=0$, then he chooses the $Z$ basis and if  $ID_{A,i}=1$, then he chooses the $X=\{\ket{+},\ket{-}\}$ basis to measure $S_{A,i}$.
            \item For $1 \leq i \leq k$, Alice and Bob construct an $k$-bit string $info(S_A)$ such that, if $S_{A,i}=\ket{0}$ or $\ket{+}$, then $info(S_{A,i})=0$, else $info(S_{A,i})=1$.
            \item They randomly choose ${k/2}$ (approximate) positions and Alice announces the values of the corresponding bits of $info(S_A)$. Bob compares these values with his corresponding measurement results to authenticate Alice's identity. Similarly Bob announces the remaining bits of $info(S_A)$ for his identity authentication. If any of them finds intolerable error rate, then he or she aborts this protocol.
        \end{enumerate}
        
        \item Message decoding: 
        \begin{enumerate}
            \item Bob discards all the decoy photons and gets back the ordered set $Q_M$.
            \item Alice announces the random permutation which she applied on $S_M$. Bob applies the inverse permutation on $Q_M$ and gets $S_M$.
            \item He measures the qubit pairs of $S_M$ in $Z\times Z$ basis or Bell basis randomly and notes the measurement results.
            \item Bob gets all the secret message bits from the measurement results of the qubit pairs of $S_M$. The relation between the measurement results and the secret message bits are given in Table~\ref{table-1}.
            To check the integrity of the secret message, Alice and Bob publicly compare some parts of the message. 
        \end{enumerate}
        
    \end{enumerate}
    \subsection{Security analysis of the modified protocol}
    We now show that our modified protocol is secure against some common attacks. First, we discuss the intercept-and-resend attack and the impersonation attack as the original YZCSS protocol was proven to be insecure against these two attacks. Then we also discuss Denial-of-Service attack, man-in-the-middle attack, entangle-measure attack and Trojan horse attack.
    \begin{enumerate}
        \item \textbf{Intercept-and-resend attack:} Let $Eve$ intercepts the ordered set $S'$ from the quantum channel. Since each qubit of the qubit pairs corresponding to the secret message is in random position, it is impossible for $Eve$ to find correct qubit pairs of $S_M$. At-most $Eve$ can do is to measure the qubits of $S'$ in $Z$ or $X$ basis. In that case, he does not get any useful information about the secret message, and also Alice and Bob detect him and terminate the protocol at the time of security checking (Step~\ref{mod-security} of the modified protocol). Note that, if Alice does not apply the random permutation on the qubits of $S_M$, then $Eve$ may get some information about the secret message, though in that case also Alice and Bob can detect his presence.
        
        \item \textbf{Impersonation attack:} In the YZCSS protocol, only Alice announces the exact states of the decoy photons corresponding to $ID_A$ and Bob compares them with his measurement results to check the authenticity of Alice. In the modified version, both Alice and Bob have to announce the information about the initial states of the decoy photons of $S_A$, they do not announce the exact states to keep $ID_A$ secret. If $Eve$ impersonating any one of Alice and Bob, then the other one can detect him and aborts this protocol (since the length $k$ of $ID_A$ is unknown to $Eve$, he can not calculate $\lambda$). Moreover, in this case also $Eve$ can not get any information about $M$ as the corresponding qubits of each qubit pairs of $S_M$ are at random positions in $Q_M$.
        
        \item \textbf{Denial-of-Service (DoS) attack:} The motivation of $Eve$, for adopting the DoS attack, is to tamper the secret message. Let $Eve$ captures the ordered set $S'$ and makes a certain operation to every qubit of $S'$. However, this action will be detected by the legitimate parties at the security checking procedure in Step~\ref{mod-security} and as a result, Alice and Bob terminate this protocol. Now suppose that $Eve$ makes changes in only a few qubits, then if the introduced error in Step~\ref{mod-security} is smaller than the threshold value, Alice and Bob can not detect $Eve$. In that case, it introduces a very small amount of error in the secret message, which is also negligible.
        
        \item \textbf{Man-in-the-middle attack:} When Alice sends the ordered set $S'$ to Bob, $Eve$ intercepts $S'$ and keep this with him. He prepares another set of qubits $S''$ and sends it to Bob. In this case, also Alice and Bob can realize the existence of $Eve$ and abort the protocol in Step~\ref{mod-security}.
        
        \item \textbf{Entangle-measure attack:}
        
        In order to steal partial information, $Eve$ may apply this attack. He first intercepts the qubits of the ordered set $S'$ and prepares some ancillary state $\ket{E}$, then applies an unitary $U_E$ to the joint states of qubits of $S'$ and $\ket{E}$ such that the composite system become entangled. However, the effect of the unitary operation $U_E$ on the second set of decoy photons are as follows:
        \begin{equation}
        \begin{aligned}
            U_E\ket{0}\ket{E}=\alpha_0 \ket{0}\ket{E_{00}}+ \beta_0 \ket{1}\ket{E_{01}},\\ U_E\ket{1}\ket{E}=\alpha_1 \ket{0}\ket{E_{10}}+ \beta_1 \ket{1}\ket{E_{11}}.
        \end{aligned}
        \end{equation}
        Since $U_E$ is unitary, we must have 
        \begin{equation}
        \begin{aligned}
        |\alpha_0|^2+|\beta_0|^2=1,\\
        |\alpha_1|^2+|\beta_1|^2=1,\\
        \alpha_0 \alpha_1^*+ \beta_0\beta_1^*=0
        \end{aligned}
        \end{equation}
        Thus when the decoy states are prepared in $Z$ basis, the error rate is $e=|\beta_0|^2=|\alpha_1|^2$.
        
        Further, we get
        \begin{equation}
        \begin{aligned}
            U_E\ket{+}\ket{E}=\frac{1}{\sqrt{2}}(\ket{+}\ket{E_{++}}+\ket{-}\ket{E_{+-}}),\\ U_E\ket{-}\ket{E}=\frac{1}{\sqrt{2}}(\ket{+}\ket{E_{-+}}+\ket{-}\ket{E_{--}}),
        \end{aligned}
        \end{equation}
        where
        \begin{itemize}
            \item $\ket{E_{++}}=\frac{1}{\sqrt{2}}(\alpha_0\ket{E_{00}}+ \beta_0\ket{E_{01}}+ \alpha_1\ket{E_{10}}+ \beta_1\ket{E_{11}})$,
            \item $\ket{E_{+-}}=\frac{1}{\sqrt{2}}(\alpha_0\ket{E_{00}}- \beta_0\ket{E_{01}}+ \alpha_1\ket{E_{10}}- \beta_1\ket{E_{11}})$,
            \item $\ket{E_{-+}}=\frac{1}{\sqrt{2}}(\alpha_0\ket{E_{00}}+ \beta_0\ket{E_{01}}- \alpha_1\ket{E_{10}}- \beta_1\ket{E_{11}})$,
            \item $\ket{E_{--}}=\frac{1}{\sqrt{2}}(\alpha_0\ket{E_{00}}- \beta_0\ket{E_{01}}- \alpha_1\ket{E_{10}}+ \beta_1\ket{E_{11}})$.
        \end{itemize}
        Thus when the decoy states are prepared in $X$ basis, the error rate is $1/2$. Thus from the error rate introduced by $Eve$ in the communication process, Alice and Bob detect this eavesdropping in Step~\ref{mod-security}. Furthermore, the random permutation applied on $S_M$ increases the security of the modified version and $Eve$ does not get any useful information about the secret message by measuring the ancillary states.
        \item \textbf{Trojan horse attack:} Both the YZCSS protocol and its modified version are one-way quantum communication protocols, i.e., only Alice prepares qubits and sends them to Bob. Thus these protocols have immunity to the Trojan horse attack.
    \end{enumerate}
    
    \section{Conclusion}\label{conclusion}
    In this paper, we analyze the security of QSDC protocols with authentication (YZCSS protocol) and demonstrate that this protocol is vulnerable to two specific attacks, namely, intercept-and-resend attack and impersonation attack. An eavesdropper adopting any one of these two attacks gets the whole secret message. The authentication process in the YZCSS protocol is unidirectional, which causes the impersonation attack. To address these concerns, we propose a modification of the YZCSS protocol, where a mutual authentication process is suggested, and the modified protocol resists the intercept-and-resend attack. We also prove that it is secure against several familiar attack strategies.
    
    \section*{Acknowledgement}
    The first author would like to acknowledge Ritajit Majumdar of Advanced Computing \& Microelectronics Unit, Indian Statistical Institute for the stimulating discussions and his insightful comments.

\bibliographystyle{unsrt}

\end{document}